\documentclass[twocolumn,showpacs,preprintnumbers,amsmath,floatfix]{revtex4}
\usepackage{graphicx}
\usepackage{dcolumn}
\usepackage{bm}

\begin{document}

\title{Comparative study of radio pulses from simulated hadron-,
electron-, and neutrino-initiated showers in ice in the GeV-PeV range}

\author{Shahid Hussain}
 \email{vacuum@ku.edu}
\author{Douglas W. McKay}%
 \email{mckay@kuark.phsx.ku.edu}
\affiliation{Department of Physics \& Astronomy 
University of Kansas, Lawrence, KS 66045.}

\date{\today}

\begin{abstract}

High energy particle showers produce coherent Cherenkov radio emission
in dense, radio-transparent media such as cold ice.  Using PYTHIA and
GEANT simulation tools, we make a comparative study among
electromagnetic (EM) and hadronic showers initiated by single
particles and neutrino showers initiated by multiple particles
produced at the neutrino-nucleon event vertex.  We include all the
physics processes and do a complete 3-D simulation up to 100 TeV for
all showers and to 1 PeV for electron and neutrino induced showers. We
calculate the radio pulses for energies between 100 GeV and 1 PeV and
find hadron showers, and consequently neutrino showers, are not as
efficient below 1 PeV at producing radio pulses as the electromagnetic
showers.  The agreement improves as energy increases, however, and by
a PeV and above the difference disappears. By looking at the 3-D
structure of the showers in time, we show that the hadronic showers
are not as compact as the EM showers and hence the radiation is not as
coherent as EM shower emission at the same frequency. We show that the
ratio of emitted pulse strength to shower tracklength is a function
only of a single, coherence parameter, independent of species and
energy of initiating particle.

\end{abstract}


\maketitle
\section{Introduction}

Ultrahigh energy (UHE) neutrinos, neutrinos with energies on the order of a
PeV and above, are predicted at some flux level by most models of the
highest energy astrophysical processes observed in the Universe. Moreover,
observations of cosmic rays above energies where gamma-nucleon thresholds
open up imply the existence of cosmogenic, or Greisen-Zatsepin-Kuzmin (GZK),
UHE neutrinos\cite{fwsgzk}. Ultrahigh energy neutrino detection will probe
interactions at energies well beyond the reach of man-made accelerators,
while bringing us directional and internal information about the
fantastically energetic sources.

A standard method of UHE neutrino detection relies on showers initiated by
the electrons and hadrons produced in neutrino-nucleon interactions. Unlike
muons and taus, electrons and hadrons dump their energy very quickly in
matter through electromagnetic and strong interactions. These showers
produce radio and optical signals \cite{askaryan1}, can be detected in
detectors like RICE \cite{ketal1,ketal2} and AMANDA \cite{amanda}, and by
those near deployment like ICECUBE \cite{icecube}, under development like
ANITA \cite{anita}, or in outline stage like SALSA \cite{salsa} or X-RICE 
\cite{xrice}. The high degree of transparency of cold ice and salt to radio
transmission make Antarctica or large salt domes attractive for radio
detection schemes like RICE, ANITA or SALSA. This paper extends earlier
studies on the signal production and propagation in cold ice. It is readily
adaptable to salt, as well.

Though considerable work has been done on shower simulations of single
particles in ice in different approximations at high energies \cite
{zhs,amz,retal}, most of the effort has focussed on EM showers and the
pulses they produce. The initial flux limit from the RICE group, for example 
\cite{ketal1}, reported an upper limit on $\nu _{e}$ flux alone. This was
conservative, in the sense that only the electron shower was used in
calculating the pulse from a charged current (CC) interaction, without the
hadronic component and with no neutral current (NC) interactions of $\nu _{e}
$ included. No $\nu _{\mu }$ or $\nu _{\tau }$ interactions were treated in
the main analysis, though estimates of hadronic contributions were discussed
and presented in an auxiliary table. More recently the preliminary results
of the analysis on an expanded subset of data \cite{ketal2} were reported,
in which the hadronic contributions to the pulses from $\nu_{e}$ CC were
estimated and included in a preliminary, updated limit. Detailed hadronic
shower and pulse studies and determination of the degree to which hadronic
and electromagnetic components are coherent were not available in that
study. Toward filling this gap we report here: (i) Full 3-D simulations of
EM showers to a PeV and hadronic showers to 90 TeV for single hadrons. (ii)
Direct, full simulation of the showers initiated by charged current
interactions of all neutrino flavors, up to a PeV. (iii) Time structure
analysis of the EM, hadronic, and neutrino showers and discussion of the
sources of difference between their pulse strengths. (iv) A universal
relationship between track length and emitted pulse strength, independent of
shower type or energy.

\section{Overview of PYTHIA/GEANT Simulations and Calculation of Signal
Strength}

For present purposes, we take our target and detection volume to be a
kilometers-cubed volume of cold ice, with an instrumented volume that is a
small fraction of this. Operationally, the calculations break into two
stages: (1) neutrino interactions with hadronic targets to produce a lepton
or neutrino plus multi-hadron final state (with PYTHIA \cite{pyth}) (2) an
electromagnetic (EM) and hadronic shower (from $\nu _{e}$ charged current
(CC) events) or only hadronic shower (from all neutral current (NC) events
and $\nu _{\mu }$ and $\nu _{\tau }$ CC events \footnote{%
We do not include the rare case of catastrophic bremstrahlung events from
final state muons in the $\nu _{\mu }$ CC interaction, and the $\tau $-decay
in the $\nu _{\tau }$ CC interaction. The $\tau $ decay length above 10 PeV
is a kilometer or greater.}) (with GEANT \cite{geant}) and the real time
calculation of the EM pulse generated by the shower particles. The output
field can be described as a pulse of field a few nanoseconds long moving out
from the interaction point on a Cherenkov cone of half angle 56$^{0}$ and
Gaussian width of several degrees \cite{zhs}. The width of the cone is
frequency dependent and, above the Landau-Pomeranchuck-Migdal (LPM) effect
threshold, energy dependent \cite{amz}. The LPM threshold is above the
energies we cover here.

To explore hadronic shower-generated pulses in detail, we expand
considerably on earlier GEANT- based shower and pulse studies \cite{retal}.
We use Geant4.5.2 to push EM\ shower analysis to 1 PeV and also include the
same analysis of $\pi ^{+}$, $\pi ^{-}$, and proton hadron-induced showers
near 100 TeV. Beyond 100 TeV, the energy limitations on some of the input
hadron cross sections prevent full development of the shower with GEANT4.5.2 
\footnote{%
We are pushing at and beyond the boundaries of directly tested physics
simulations in our PYTHIA and GEANT applications. The results we quote are
reasonable indicators of the physics above a TeV, but far from the last word}%
.

To simulate neutrino induced showers and the radio frequency pulses they
produce, we use PYTHIA6.205 to simulate neutrino-nucleon CC interaction. The
PYTHIA output data files are fed to GEANT and relevant shower and field
parameters calculated as the shower simulation proceeds. The Fourier
transform of the electric field in the Fraunhoffer limit is calculated for
each track segment of every particle in the shower and the contributions of
the track segments to the total field are summed vectorially, as described
in \cite{zhs}. We follow the standard practice and report the magnitude of
the field at a given angle and frequency multiplied by the distance from the
start position of the shower to the position of the observer. This value is
a function of frequency, angular position of the observer and energy of the
initiating particle. We generally refer to this as ``the pulse''. The phase
of the field produced by each track segment depends upon the track start and
stop times and initial track position, which feeds local information into
the total pulse.

We report here the essential features of showers and the pulses they
generate. We want to find out whether the features of hadron induced showers
and pulses at a given energy are similar to EM features at the same energy,
whether the comparison is energy dependent, and whether the multi-hadron
final states from $\nu _{e}$ CC interactions produce pulses that are
essentially coherent with the pulse the electron produces. We consider
electron and $\nu _{e}$ CC induced events first and then turn to hadron and $%
\nu _{\mu }$ and $\nu _{\tau }$ CC-induced events. The simulation results of
CC-induced events of $\nu _{\mu }$ and $\nu _{\tau }$ can also be used to
estimate the results for NC-induced events as both are initiated
hadronically (tau and muon decay lengths are large as compared to the
typical shower depths at high energies. Hence, for a given value of
inelasticity, we expect $\nu _{\mu }$ and $\nu _{\tau }$ CC-induced showers
to be scalable to the NC-induced events with the same inelasticity value).

The simulations were done in 3-D, including all the hadronic and EM physics
processes involved. For each particle except neutrinos, we have simulated
100 showers at 100 GeV, 50 showers at 1 TeV, 10 showers at 10 TeV, single
showers at 90/100 TeV and 1 PeV. For the neutrinos we have simulated 100
showers at 1 TeV, 30 showers at 10 TeV, and 5 showers at 100 TeV. The mean
deviations of the non-EM numbers given in Table~\ref{tab:showchar} are in
the range 1-8\%; although we cannot directly determine the error in the data
at 90 TeV and above, as we have only single showers, we expect it will
certainly be below 10\% because a single shower will have many times the
number of particles in a 10 TeV shower, so the relative fluctuations will be
smaller in a shower around 100 TeV and above.

\begin{table}[tbp]
\caption{Different characteristics of the showers as a function of energy:
1st column is energy in TeV; 2nd column has the particle name (for neutrinos
the elasticity value is 0.65 in every case) ; $\Delta
r_{z}=r_{z}^{-}-r_{z}^{+}$ in the 3rd column gives the difference of the
track lengths in meters, for positive and negative charges, projected along
z direction (i.e. direction of propagation of the primary particle); 4th
column ($R|E|1$) gives $R|E|$ (see text for definition of $R|E|$) values in $%
\mu V/MHz$ at the Cherenkov angle, at 1 GHz (see Section II for the
estimated random errors of these numbers); 5th column gives $w = \frac{%
\Delta r_{z}^{c}}{\Delta r_{z}}\ $ (its meaning is discussed in section
III); 6th column gives $b(w)*10^{5}$ (see Eq. 2); 7th column gives $R|E|2$
which is the pulse value calculated from Eq. 2.}
\label{tab:showchar}
\begin{tabular}{|l|l|l|l|l|l|l|}
\hline
$E$ & $name$ & $\Delta r_{z}$ & $R|E|1$ & $w$ & $b(w)*1e5$ & $R|E|2$ \\ 
\hline
$
\begin{array}{l}
0.1
\end{array}
$ & $
\begin{array}{l}
e^{-} \\ 
\pi ^{-} \\ 
\pi ^{+} \\ 
p
\end{array}
$ & $
\begin{array}{l}
136 \\ 
70 \\ 
68 \\ 
53
\end{array}
$ & $
\begin{array}{l}
0.014 \\ 
0.0034 \\ 
0.0034 \\ 
0.0026
\end{array}
$ & $
\begin{array}{l}
0.90 \\ 
0.62 \\ 
0.62 \\ 
0.56
\end{array}
$ & $
\begin{array}{l}
9.9 \\ 
4.7 \\ 
4.7 \\ 
4.8
\end{array}
$ & $
\begin{array}{l}
0.013 \\ 
0.0033 \\ 
0.0032 \\ 
0.0026
\end{array}
$ \\ \hline
$
\begin{array}{l}
1
\end{array}
$ & $
\begin{array}{l}
e^{-} \\ 
\pi ^{-} \\ 
\pi ^{+} \\ 
p \\ 
\nu _{e} \\ 
\nu _{\mu } \\ 
\nu _{\tau }
\end{array}
$ & $
\begin{array}{l}
1336 \\ 
898 \\ 
897 \\ 
812 \\ 
1156 \\ 
269 \\ 
271
\end{array}
$ & $
\begin{array}{l}
0.13 \\ 
0.058 \\ 
0.057 \\ 
0.046 \\ 
0.087 \\ 
0.012 \\ 
0.012
\end{array}
$ & $
\begin{array}{l}
0.90 \\ 
0.75 \\ 
0.74 \\ 
0.71 \\ 
0.81 \\ 
0.60 \\ 
0.60
\end{array}
$ & $
\begin{array}{l}
9.9 \\ 
6.4 \\ 
6.3 \\ 
5.7 \\ 
7.8 \\ 
4.6 \\ 
4.6
\end{array}
$ & $
\begin{array}{l}
0.13 \\ 
0.058 \\ 
0.056 \\ 
0.046 \\ 
0.091 \\ 
0.012 \\ 
0.012
\end{array}
$ \\ \hline
$
\begin{array}{l}
10
\end{array}
$ & $
\begin{array}{l}
e^{-} \\ 
\pi ^{-} \\ 
\pi ^{+} \\ 
p \\ 
\nu _{e} \\ 
\nu _{\mu } \\ 
\nu _{\tau }
\end{array}
$ & $
\begin{array}{l}
13319 \\ 
9963 \\ 
9728 \\ 
9210 \\ 
12095 \\ 
3296 \\ 
3395
\end{array}
$ & $
\begin{array}{l}
1.3 \\ 
0.75 \\ 
0.70 \\ 
0.64 \\ 
1.1 \\ 
0.22 \\ 
0.25
\end{array}
$ & $
\begin{array}{l}
0.90 \\ 
0.79 \\ 
0.78 \\ 
0.77 \\ 
0.86 \\ 
0.77 \\ 
0.78
\end{array}
$ & $
\begin{array}{l}
9.9 \\ 
7.4 \\ 
7.1 \\ 
6.9 \\ 
9.0 \\ 
6.9 \\ 
7.1
\end{array}
$ & $
\begin{array}{l}
1.3 \\ 
0.73 \\ 
0.69 \\ 
0.64 \\ 
1.1 \\ 
0.23 \\ 
0.24
\end{array}
$ \\ \hline
$
\begin{array}{l}
10^{2} \\ 
90 \\ 
10^{2} \\ 
90 \\ 
10^{2} \\ 
10^{2} \\ 
10^{2}
\end{array}
$ & $
\begin{array}{l}
e^{-} \\ 
\pi ^{-} \\ 
\pi ^{+} \\ 
p \\ 
\nu _{e} \\ 
\nu _{\mu } \\ 
\nu _{\tau }
\end{array}
$ & $
\begin{array}{l}
1.33E5 \\ 
1.04E5 \\ 
1.19E5 \\ 
90007 \\ 
1.26E5 \\ 
39982 \\ 
36800
\end{array}
$ & $
\begin{array}{l}
13 \\ 
9.2 \\ 
11 \\ 
6.9 \\ 
12 \\ 
3.5 \\ 
3.0
\end{array}
$ & $
\begin{array}{l}
0.90 \\ 
0.85 \\ 
0.86 \\ 
0.80 \\ 
0.87 \\ 
0.84 \\ 
0.82
\end{array}
$ & $
\begin{array}{l}
9.9 \\ 
8.8 \\ 
9.0 \\ 
7.5 \\ 
9.3 \\ 
8.5 \\ 
8.0
\end{array}
$ & $
\begin{array}{l}
13 \\ 
9.1 \\ 
11 \\ 
6.8 \\ 
12 \\ 
3.4 \\ 
3.0
\end{array}
$ \\ \hline
$
\begin{array}{l}
10^{3}
\end{array}
$ & $
\begin{array}{l}
e^{-} \\ 
\nu _{e} \\ 
\nu _{\mu } \\ 
\nu _{\tau }
\end{array}
$ & $
\begin{array}{l}
1.33E6 \\ 
1.32E6 \\ 
4.08E5 \\ 
3.67E5
\end{array}
$ & $
\begin{array}{l}
132 \\ 
131 \\ 
37 \\ 
30
\end{array}
$ & $
\begin{array}{l}
0.90 \\ 
0.89 \\ 
0.86 \\ 
0.82
\end{array}
$ & $
\begin{array}{l}
9.9 \\ 
9.7 \\ 
9.0 \\ 
8.1
\end{array}
$ & $
\begin{array}{l}
132 \\ 
128 \\ 
37 \\ 
30
\end{array}
$ \\ \hline
\end{tabular}
\end{table}
\begin{figure}[tbp]
\includegraphics[width=3.2in,angle=0]{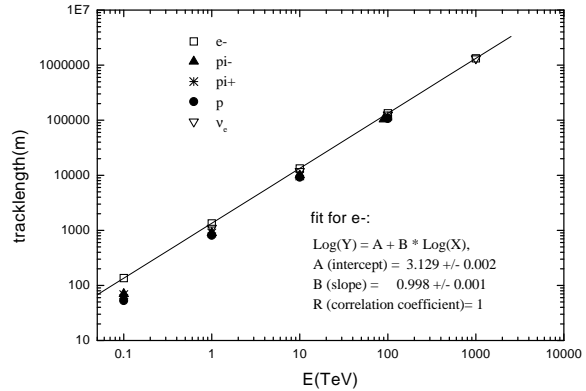}
\caption{Projected weighted-track length difference $r_{z}^{-}-r_{z}^{+}$
(see text for more detail) vs energy for different particles. Equation of
the linear fit for $e^{-}$ is also shown. For the neutrino case, elasticity (%
$1-y$) is 0.65+/-0.01.}
\label{fig:fig1}
\end{figure}

\begin{figure}[tbp]
\includegraphics[width=3.2in,angle=0]{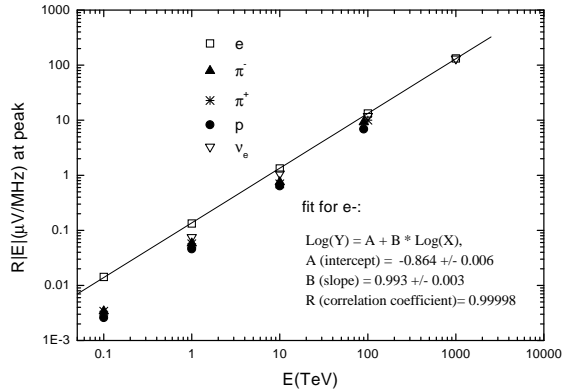}
\caption{$R|E|$ at $\theta _{c}$ and 1 GHz vs energy for different
particles. Equation of the linear fit for $e^{-}$ is also shown. For the
neutrino case, the elasticity is the same as in figure~\ref{fig:fig1}.}
\label{fig:fig2}
\end{figure}

\subsection{Electron and Electron-Neutrino Charged Current Induced Pulses}

As in earlier work all field calculations are done in the Fraunhoffer limit $%
R\gg \frac{L^{2}}{\lambda }$, where $R$ is detector-shower distance, $%
\lambda $ is the wavelength, and $L$ is the shower penetration distance. The
simulation is run in cold ice with $\rho =0.92g/cm^{3}$ , and with $A=14.3$, 
$Z=7.2$, and radiation length=38.8cm, as calculated by GEANT. 

\begin{figure}[tbp]
\includegraphics[width=3.2in,angle=0]{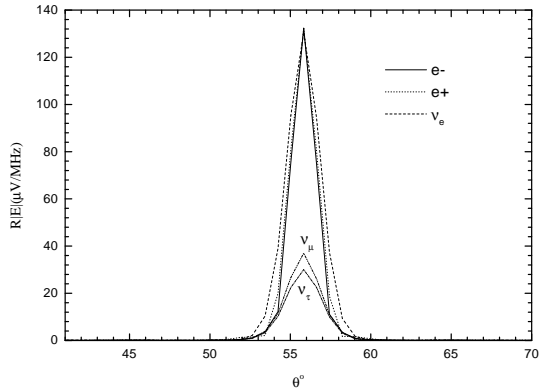}
\caption{$R|E|$ vs $\theta$, at 1 GHz, and E=1 PeV for single shower
simulations. For neutrinos, the elasticity is the same as in figure~\ref
{fig:fig1}.}
\label{fig:fig3}
\end{figure}

Injecting electrons at 0.1, 1, 10, 100, and 1000 TeV, we calculate $\Delta
r_{z}=r_{z}^{-}-r_{z}^{+}$, the total weighted track lengths projected along
shower axis, shower charged particle profiles, frequency dependence of the
field ($|E|$) times the distance R to the detector (the field strength
decreases as 1/R in the Fraunhoffer limit), and angular dependence of the
field at 1 GHz. The results for key parameters are shown in Table~\ref
{tab:showchar} . The electron is the first entry at each energy. As is
apparent from the numbers, the projected track length difference ($\Delta
r_{z}$) between the negative and positive particles and the pulse height at
a fixed frequency and at the Cherenkov angle grow linearly with energy to
high accuracy. This is shown in Figs.~\ref{fig:fig1} and ~\ref{fig:fig2},
where the open boxes are the electron data and the fits are shown on the
graph. These results agree with all earlier work \cite{zhs,retal}. The
angular distribution is shown in Fig.~\ref{fig:fig3} for $E=1 PeV$ in the
region of the Cherenkov angle. This plot clearly shows the rapid suppression
of signal off the Cherenkov cone. Injecting 1, 10, 100, and 1000 TeV
neutrino-nucleon interaction output from PYTHIA, we calculate the
corresponding quantities with elasticity $1-y=0.65 \pm 0.01$ at each energy. 
We choose 
the same $y$ at each energy so that we can determine whether the hadronic component of the 
entire shower becomes more efficient at generating pulse strength with increasing energy. 
This requires keeping 
the hadronic fraction of the total energy constant as we change energy. The choice $y=0.35$ 
turns out to be big enough so that hadronic fluctuations from shower-to-shower are 
reasonably small and at the same time the electron does not completely dominate the 
signal in the $\nu_e$ case. The trends in the behaviour with neutrino energy that we 
describe below are true regardless of the $y$-value chosen. 

In Table~\ref{tab:showchar} , selected examples of these calculations are given at each
energy. As is well known \cite{gqrs}, the elasticity peaks at 1 and the
average grows slowly with energy, starting at ???? at 0.1 TeV and 
reaching about 0.74 at $E_{\nu }=1 PeV$. 
In incident neutrinos of given flavor and energy, we assembled a sample within our 
chosen range of $y$ by generating PYTHIA events and selecting those in our chosen 
$y$ range until our sample size was reached. As mentioned above, our aim is to 
study the pulse-production efficiency of the hadronic component of neutrino showers 
as a function of energy. In other words, is the pulse growth slowers than, the same 
as, or faster than the growth in hadronic energy? We need the fraction ($y$), of 
hadronic energy in the shower to be held constant as we raise the energy to 
answer this question. Since average $y$ decreases as the energy increases, the 
average $y$ at a given energy is not the best choice for our purpose here. For application 
to simulation of signals for an actual experiment, one would {\it not} use fixed 
$y$ of course. 
At a given $y$ value, the fluctuations in shower and pulse parameters
decrease with increasing energy and at the same time the $\nu _{e}$ results
look more and more like electron results; the $\nu _{e}$ event at 1 PeV
gives projected, weighted track length and pulse values that are essentially
the same as those of the electron event. This is clearly shown in Fig.~\ref
{fig:fig3} as well. 

The trend of our simulation data shows that, as the
primary energy increases, the hadronic component of the events {\it for any $y$
value} contributes in such a way that the $\nu _{e}$ events look more and
more like electron events as the primary energy increases. We will see this
in a more direct way in the results from single injected hadron, which we
describe in the next subsection.

\subsection{Hadron, and $\nu _{\mu }$ and $\nu _{\tau }$ CC Induced Pulses}

The calculation of the charged pion and proton induced events proceed
similarly to those of the electron/positron induced events. There are many
more processes included in the hadronic cases, of course; the maximum energy
of pion and proton showers is around 100 TeV in GEANT4.5.2, because some
cross sections, such as the kaon- total cross sections, are extrapolated
only to about 10 TeV. Nonetheless, we can gain valuable insight from the
comparison between electron and hadron-induced events. Since the low energy
tail of a shower is essentially EM, and the excess charge develops in this
low energy region, the hadron shower should behave similarly to the EM
showers to the extent that gammas from $\pi ^{o}$ decay dominate the end of
the shower. This should be a better picture as energy increases and
fluctuations in particle species wash out.

The essential features of the comparisons between the electron and
hadron-induced showers are summarized in Figs.~\ref{fig:fig1} and ~\ref
{fig:fig2}. The projected weighted track length and pulse height at
Cherenkov angle at fixed frequency grow linearly with energy. It is clear
from the graphs (Figs.~\ref{fig:fig1} and ~\ref{fig:fig2}) and from a
look at the numbers themselves (Table~\ref{tab:showchar}) that the
efficiency of converting hadron input energy increases with energy, as we
emphasize further below. It is important to note that a given weighted
projected track length in a hadronic shower is not as efficient in producing
the radio signal as is the same length of EM shower. However, this
difference between the hadronic and the EM showers shrinks at higher
energies. As we will discuss in the next section, this is because of the
difference between hadronic and EM showers in the distribution in space of
the particles contributing to the weighted, projected track length.

As we see in Fig.~\ref{fig:fig2}, the pulse outputs from all the showers
are nearly equal as one approaches 1 PeV shower energy. We expect that the
hadron curve will flatten out and follow the electron at higher energies,
though we cannot verify this with the current GEANT capabilities. The
tendency for hadron events to act more like EM events is consistent with our
observation earlier that the $\nu _{e}$ events act more and more like
electron events as we raise the injection energy; the hadron component adds
coherently with the EM component.

The $\nu _{\mu }$ and $\nu _{\tau }$ CC results are the final entries at
each energy. Here again it is clear that the hadronic component of the final
state is becoming more efficient in converting energy to EM pulse strength
as the injection energy increases. For example, the average $\nu _{\mu }$ or 
$\nu _{\tau }$ event at 10 TeV with our illustrative value $1-y=0.65$ 
produces a significantly
larger fraction of the electron pulse than it does at 1 TeV. The
corresponding comparison at 1 PeV for $\nu _{\mu }$ and $\nu _{\tau }$
events produce noticeably different shower and pulse results at the same
energy and elasticity. We looked at the detailed structure of the PYTHIA
output for these and lower energy cases and found that the larger track
length and larger pulse result correspond to the event with larger $\pi ^{o} 
$ content in the PYTHIA output (GEANT4 input) (and, ultimately, higher
shower EM-content; this explains why $\nu _{\mu } $ produces a stronger
pulse than $\nu _{\tau }$ with the same E and $y$ in one example and weaker
in another. Even at 1 PeV, there are significant fluctuations in the species
and subenergy compositions of the $\nu +p\rightarrow l+X$ final states and
these fluctuations produce corresponding shower and pulse fluctuations. The
trends in Figs.~\ref{fig:fig1} and ~\ref{fig:fig2} clearly indicate that
above 1 PeV these fluctuations damp out and hadronic components of the
showers act like electromagnetic showers of the same energy. 
The LPM effect is expected to set
in at higher energies and elongate the EM showers but not the hadronic
showers \cite{amz}.

We have also calculated the frequency spectrum of the pulses at Cherenkov
angle from each type of shower at each energy. In summary, we find the
spectra from different types of shower look more and more similar to each
other as the shower energies increase. By 10 TeV, they all rise sharply from
zero at zero frequency, peak at several GHz, fall and then level out in the
10 to 20 GHz region and are essentially flat above that. The frequency
behavior up to 1 TeV is discussed for electron showers in S. Razzaque et al. 
\cite{retal}, where, as in our present study, the sharp turn over of the
spectrum does not set in until 1 TeV. We find that the hadron and hadron
dominated $\nu_{\mu}$ and $\nu_{\tau}$ do not show this behavior until and
order of magnitude higher energy.

\section{Time structure analysis of hadronic and EM showers}

As far as we are aware, direct time development of EM, hadronic and
neutrino-induced showers in ice has not been reported. Here we describe
several analyses to clarify the reasons behind the generally improved
efficiency in radio pulse production by hadronic showers of increasing
energy. We focus on several direct-time evolution representatives of the
shower content for this purpose. The weighted, projected track length $%
\Delta r_{z}$ =$r_{z}^{-}-r_{z}^{+}$, described earlier, is known to be a
good indicator of pulse height produced by the EM showers \cite{zhs}. As
Table~\ref{tab:showchar} shows, this is not the whole story in understanding the difference
between the pulses produced by electron showers vs. hadron showers at lower
energies. For example, at 1 TeV, the ratios of $r_{z}^{-}-r_{z}^{+}$ for $%
e:\pi ^{-}$, $e:\pi ^{+}$, and $e:p$ are 1.49, 1.49, and 1.65, respectively,
while the corresponding $R|E|$ ratios are 2.24, 2.48, and 2.83. The
corresponding comparisons for $\nu_{e}$, $\nu_{\mu}$ and $\nu_{\tau}$ are
1.16, 4.95 and 4.95 compared to 1.53, 11.0 and 11.0. The ratio of pulse to
track length is larger in the EM than in the hadronic showers and, to the
extent they are dominated by hadrons, the neutrino showers. The EM showers
are evidently producing more coherence in the Cherenkov emission.
Intuitively one expects the degree to which the emission is coherent depends
upon the degree to which the charged particles, and therefore the charge
imbalance, are concentrated in a small volume, acting like a single, point
charge.

To get a visualization of these features as the shower develops, in Figs. 4
and 5 we plot the X, Y, and Z components of the charged particle position
versus time at each step in every track's development in a 100 GeV electron
and proton showers. The lightest shaded points are X, next darker Y, and the
darkest is the Z component of all charged particle positions as a function
of time. Beyond 50 ns, there are no tracks in electron showers and only a
tiny percentage in hadron showers. The speed-of-light line in cold ice is
designated by a heavy dark line. A comparison between the plots vividly
shows that the X and Y excursions of electron shower particles are smaller
than those of the proton shower of the same energy. The concentration of the
Z components near the speed of light in vacuum is correspondingly higher
than that in the proton shower. These plots indicate that the coherence of
the emission from the electron shower should be greater than the one from a
proton shower, which correlates well with the fact that the emission is
weaker from the proton shower at 1 GHz than is expected just from the $%
r_{z}^{-}-r_{z}^{+}$ comparison.

Finally we examined the degree to which the value of $r_{z}^{-}-r_{z}^{+}$
is determined by the particles which are highly relativistic and are hugging
a point moving with the speed of light along Z-axis. For energies of 0.1
TeV, 1 TeV, 10 TeV and 90 TeV \footnote{%
At energies around 100 TeV and above, hadronic shower runs crash because at
least one daughter kaon energy is above the maximum simulated in the current
version of Geant.}, we compute the value of $r_{z}^{-}-r_{z}^{+}$ for just
those tracks with points within a distance 
\begin{equation}
r_{core}=\sqrt{X^{2}+Y^{2}+(Z-ct)^{2}}\leq 0.15m,  \label{core}
\end{equation}
as shown in Fig. 6 (only two dimensions are shown). The choice a = 0.15m is
guided by
the wavelength at 1 GHz. This choice, though not unique by
any means, is satisfactory to make our point about the role of compactness
and coherence \footnote{%
For example the radial distribution of net charge for the whole shower
suggests a value smaller than 0.15m, as shown in Razzaque et al.\cite{zhs}}.
The idea is simply that charge dispersed farther from the shower core will
tend to degrade coherence. We have calculated the projected weighted-track
lengths $\Delta r_{z}^{a}=r_{z}^{-}-r_{z}^{+}$ of the particles within the
shaded region of the sphere with radius a=0.15m (see figure 6) and the
projected weighted-track lengths $\Delta r_{z}=r_{z}^{-}-r_{z}^{+}$ due to
all the charged particles in the shower.

In Table~\ref{tab:showchar}  we show the ratio $w=\frac{\Delta r_{z}^{a}}{\Delta r_{z}}$ for
the showers initiated by electrons, pions, protons and the three flavors of
neutrinos at various energies. As we see here, this ratio has a constant
value of 0.90 for the electron initiated showers between energies 100 GeV-1
PeV. This means 90\% of the projected weighted-track length comes from a
compact region defined by Eq. \ref{core}. On the other hand this ratio is
smaller for a proton, a pion, or a neutrino shower and it improves as we go
to higher energies. For EM showers, the value of $\Delta r_{z}$ has long
been recognized as an extremely reliable measure of the pulse height
produced at Cherenkov angle for frequencies at and below 1 GHz. We emphasize
here the added insight gained by treating separately the track length of the
core region defined in Eq. 1, and the track length outside this region. If
we look at Table~\ref{tab:showchar}  and the Figs. 2 and 3, we see as the ratio $w$ for
hadronic showers approaches that of the electron shower (Table~\ref{tab:showchar} ), the
weighted, projected track lengths and pulse heights of the hadronic showers
also approach those of the electron showers. The well established linear
relationship between electron shower projected track length and energy and
between pulse height and energy is expressed in terms of our ``compactness''
parameter as a constant value $w=0.90$ for all electron showers between 0.1
TeV and 1.0 PeV.

As is clear from the numerical results summarized in Table~\ref{tab:showchar}  and our
discussion above, the pulse strength at the Cherenkov angle at a given
frequency is related to the weighted, projected track length, $\Delta r_{z}$%
, and the ratio $w$. The data suggest that the ratio of $R|E|$ to $\Delta
r_{z}$ is a function of $w$ that is independent of the particle type and
energy. For example, there are three entries with $w=0.86$: $\nu _{e}$ at 10
TeV, $\pi ^{+}$ at 100 GeV, and $\nu _{\mu }$ at 1 PeV; these have values $%
R|E|/\Delta r_{z}=9.1$ $10^{-5}$, 9.2 $10^{-5}$ and 9.1 $10{^{-5}}$,
respectively. This and similar results in Table~\ref{tab:showchar} indicate 
there is a relationship

\begin{equation}
R|E|/\Delta r_{z}=b(w),
\end{equation}
where $w$ is defined above and one determines the function $b(w)$ from the
data. We find that a third degree polynomial in $w$ fitted to the data
plotted as $R|E|/\Delta r_{z}$ vs. $w$ gives a remarkably good description
of the data in Table~\ref{tab:showchar} \footnote{%
The polynomial is $b(w) = c_{0} + c_{1}w + c_{2}w^2 + c_{3}w^3$, with $c_{0}
= 7.85e-4, c_{1} = 30.4e-4, c_{2} = 39.8e-4$, and $c_{3} =16.1e-4 $. To fit
the data above 1 TeV, a good quality \textit{linear} fit can be found}. Given
the value of $w$ and $\Delta r_{z}$, one reproduces the value of $R|E|$
within a few percent for all but a couple of the 29 entries, and those two
are good to better than about $6\%$. As remarked in footnote 21, above 1
TeV, a \textit{linear} fit works quite well. This universal feature supports
our picture that the key factor that relates the pulse at a given frequency
to the weighted, projected track length for showers induced by any particle
in the energy range we consider is the degree of coherence of the emission,
as determined by the compactness of the relativistic core of the shower as
it moves along through the medium.

\section{Summary and Conclusions}

Ultrahigh energy neutrinos will provide a uniquely clear window to explore
fundamental physics and astrophysics processes at the highest known
energies. These neutrinos produce huge particle showers when they interact
with dense matter. The showers then produce coherent, Cherenkov radio
signals, the strength of which depends on the net charge in the shower, on
the track length projected along the shower axis and on the degree of
compactness of the shower. All of these factors increase with neutrino
energy. Combined with a radio-transparent medium such as cold ice or pure
salt, the radio detection technique becomes increasingly powerful as energy
rises in the UHE regime.

The particle content early in a shower depends on the flavor of the neutrino
producing it. Electron-neutrino showers immediately have a heavily
electromagnetic character, which dominates the hadronic component on
average. On the other hand, the muon-neutrino and taon-neutrino showers are
essentially hadronic in character. Below a PeV our simulation results show
that hadronic showers are not as efficient in producing radio signals as EM
showers. Though the pulse strengths produced by hadron showers increase with
energy, the net charge, the weighted, projected track length, and the
corresponding pulse strength are all smaller than those of the EM showers at
the same energy. In addition, we found that the pulse strength per weighted,
projected track length is smaller for the hadronic showers than for the
electromagnetic ones. However, the key result of our study is that the
differences in the efficiencies of pulse production disappear as one goes to
higher and higher energy. The same is true of the comparisons between the
neutrino showers and the electron showers. Extrapolating hadronic shower
results to 1 PeV from our highest simulation results at 100 TeV, we find
that the hadron and electron showers at a PeV give essentially the same
pulse strengths. Moreover, at 1 PeV, the simulated $\nu_{e}$ shower with elasticity of
0.65 is indistinguishable from the electron shower at the same energy. The $%
\nu_{\mu}$ and $\nu_{\tau}$ showers continue to show convergence toward
electron shower behaviour when the elasticity factor in the events is
accounted for.

In a new type of study employing time- and space- structure of the showers,
we showed that the low energy electron showers are significantly more
compact than the corresponding hadronic showers and neutrino showers at low
energies. The coherence is directly related to the compactness and therefore
to the efficiency of converting charge imbalance and track length into EM
pulses. We show that a simple, universal formula nicely describes this
aspect of the simulation data. Again, as one goes to ultrahigh energies,
hadronic and EM showers become equally compact and equally coherent and,
consequently, equally efficient in producing radio signals . All the signals
rises linearly with energy at ultrahigh energies and with essentially the
same normalization. We believe our study develops a detailed account not
previously available of how his simple picture emerges in the 100 GeV to 1
PeV energy regime.

We conclude that at and above a PeV, until the Landau-Pomeranchuk-Migdal
(LPM) effect becomes important\cite{rrj}, the hadronic and electromagnetic
showers produce essentially the same pulses at a given energy. In addition,
the hadronic and electromagnetic components of the $\nu_{e}$ pulses are
effectively coherent. Because the LPM effect has bearing on the EM particles
that are above LPM threshold, and the EM particles that evolve from UHE
hadronic showers up to EeV energies are below this threshold \cite{amz}, the
hadronic showers are predicted to become more effective than the EM showers
at producing EM pulses at some point above the LPM threshold. Further study
of this important problem is certainly needed and will be the subject of
future research.

\section*{Acknowledgments}

We benefitted greatly from discussions and communications with Dave Besson,
Will Chambers, John Ralston, Soeb Razzaque, Dave Seckel, Suruj Seunarine
during this course of this work, which was supported in part by The 
Department of Energy under grant number DE-FG03-98ER41079. Doug McKay 
thanks the staff and physics group of the Phenomenology Institute at the 
University of Wisconsin for its hospitality while this work was being completed.

\begin{center}
{\LARGE Figure Captions for Figures 3-6 (separate .jpg files)}
\end{center}

FIG. 4: 3-D structure in time of a 100GeV electron shower. The light gray
color represents X, dark gray Y, and black Z coordinates. For the Z case,
particles below the thick solid line z=c*time/1.8 are moving slower than the
speed of light in cold ice.

FIG. 5: Same as figure 4 for a 100GeV proton shower.

FIG. 6: The circle shown here has a radius a=0.15m and is centered at a
point that moves with the speed of light in vacuum. The shaded region is
defined by Eq. 1 in the text.

\end{document}